\documentclass[rsi,reprint,graphicx]{revtex4-1}
\usepackage{graphicx}
\usepackage{amsmath}

\draft 

\begin{document}


\title{Multiplexed readout of kinetic inductance bolometer arrays} 



\author{Hannu Sipola$^a$, Juho Luomahaara$^a$, Andrey Timofeev$^a$, Leif~Gr\"onberg$^a$, Anssi Rautiainen$^b$, Arttu Luukanen$^b$, and Juha~Hassel$^a$}
\email[]{Juha.Hassel@vtt.fi}
\affiliation{$^a$VTT Technical Research Centre of Finland Ltd., QTF Centre of Excellence, P.O. box 1000, FI-02044 VTT, Finland}
\affiliation{$^b$Asqella Oy, Kutomotie 18, FI-00380 Helsinki, Finland}


\date{\today}

\begin{abstract}
Kinetic inductance bolometer (KIB) technology is a candidate for passive sub-millimeter wave and terahertz imaging systems. Its benefits include scalability into large 2D arrays and operation with intermediate cryogenics in the temperature range of 5 -- 10 K. We have previously demonstrated the scalability in terms of device fabrication, optics integration, and cryogenics. In this article, we address the last missing ingredient, the readout. The concept, serial addressed frequency excitation (SAFE), is an alternative to full frequency-division multiplexing at microwave frequencies conventionally used to read out kinetic inductance detectors. We introduce the concept, and analyze the criteria of the multiplexed readout avoiding the degradation of the signal-to-noise ratio in the presence of a thermal anti-alias filter inherent to thermal detectors. We present a practical scalable realization of a readout system integrated into a prototype imager with 8712 detectors.  This is used for demonstrating the noise properties of the readout. Furthermore, we present practical detection experiments with a stand-off laboratory-scale imager.
\end{abstract}

\pacs{}

\maketitle 

\section{Introduction}

Superconducting detectors have an established position in submillimeter wave, terahertz, and far infrared band radiometric imaging systems. The state of art in the detector integration level is today defined by superconducting transition edge sensors (TESs) \cite{irw1} and kinetic inductance detectors (KIDs) \cite{day1,bas2} used for the application of astronomical imaging. Detector arrays with up to 10000 sensing elements have been demonstrated \cite{hol1,bas1}. A further application of submillimeter-wave radiometry is security screening which relies on good penetration of radiation through dielectric materials, while a typical spatial resolution is in the centimeter range as fundamentally limited by diffraction \cite{hei1,row1,luu1,luu2}. This enables the detection of concealed objects under the clothing. Large arrays arrays are beneficial in terms of system-level figures of merit such as spatial resolution, radiometric contrast, field of view, and mechanical simplicity. One limiting aspect in the scaling is the readout. For TES-based systems, established solutions include rather involved SQUID-based  readouts using either time-domain or frequency-domain multiplexing \cite{irw2}. The readout of large KID arrays using RF or microwave frequency multiplexing typically requires state-of-art high-speed digital electronics operating at gigahertz sampling rates \cite{bas1,hug1}. 

We have previously introduced the detector technology based on kinetic inductance bolometers (KIBs) \cite{tim1}. In brief, KIBs are thermal detectors based on sensing the change of the temperature-dependent kinetic inductance on a thermally confined membrane that is heated by the signal absorbed on the membrane. Compared to non-equilibrium KIDs and TESs relying in sub-Kelvin cryogenics, KIBs operate in the temperature range of 5 -- 10 K. We have previously demonstrated the scalability into kilo-pixel arrays, and performance in line with the requirements of radiometric imaging of room temperature objects \cite{tim2}. In this paper, we concentrate on a readout concept suitable for the KIB arrays.

Kinetic inductance detectors and sensors are inherently compatible with RF or microwave multiplexing, enabling separate resonance tuning of each sensor element with an individual readout frequency \cite{day1,bas2,luo1}. The readout is typically performed with frequency-division multiplexing (FDM) addressing a number of detectors in parallel with a frequency comb, usually at microwave frequencies. In these approaches, the comb generation and demodulation require fast digital electronics. Here we introduce  an alternative method, serial addressed frequency excitation (SAFE), which substantially simplifies the readout system. After introducing the concept, we discuss the criteria to avoid potential multiplexing-induced degradation of the signal-to-noise ratio (SNR), and present experimental results from the KIB-readout supporting the theory. We also describe an implementation of the electronics integrated into an imaging system, and show radiometric detection results confirming the capacity of the readout in concealed object detection.
\begin{figure*}[!t]
\begin{center}
\includegraphics[width = 18cm]{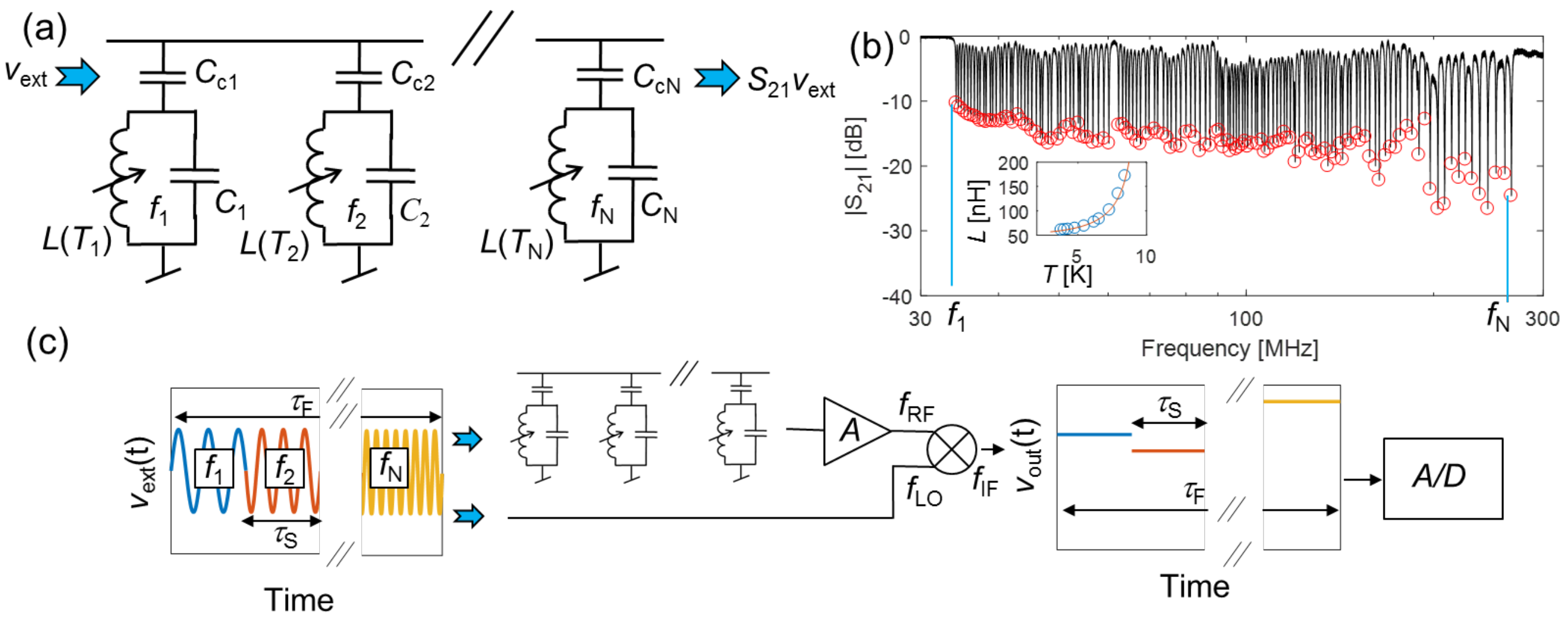}
\end{center}
\caption{(a) The readout-band electrical equivalent circuit of a KIB array corresponding to one readout channel with $N$ detectors. Each temperature-dependent inductance $L\left(T_i\right)$ is tuned by an individual shunt capacitance $C_i$, leading to characteristic resonance frequencies $f_i$. The resonators are matched to the readout line with coupling capacitors $C_{ci}$. (b) The readout band transmission amplitude $|S_{21}|$ of a single readout channel with $N=132$ detector elements as recorded with a vector network analyzer. The resonant frequencies are denoted by red circles. The inset shows the temperature dependence of the inductance $L\left(T_i\right)$ as extracted from one resonance frequency shift. (c) A conceptual illustration of SAFE. The excitation signal $v_{ext}$ has time-switching frequency addressing the readout resonances $f_i$. The excitation sequence consists of frames with duration $\tau_F$ during which each detector is addressed within a time slot of length $\tau_S = \tau_F/N$. As driven through the KIB array, the excitation tones get modulated by corresponding detector signals. The modulated signal $v_\mathrm{out} = S_{21}v_\mathrm{{ext}}$ is then amplified ($\mathrm{A}$) and fed to the radio frequency (RF) port of the mixer. By using the unmodulated excitation tone as the local oscillator (LO) the mixer intermediate frequency (IF) output port produces the demodulated signal, i.e., time-multiplexed detector signals to be analog-to-digital (A/D) converted for post processing and image formation.}
\end{figure*}

\section{The readout concept}

\subsection{Basic functionality}

A readout-band schematic of a KIB array with $N$ detectors corresponding to one readout channel is shown in Fig. 1(a). For KIBs, the temperature of the $i$:th bolometer is $T_i = T_b + \Delta T$, where $T_b$ is the bath (cryostat) temperature,  and $\Delta T$ depends on the radiometric signal. This causes a shift in the inductance $L\left(T_\mathrm{i}\right)$ which in turn shifts the resonant frequncy $f_\mathrm{i}$ of the resonator formed by $L\left(T_\mathrm{i}\right)$ and the shunt capacitance $C_\mathrm{i}$. The coupling capacitors $C_{\mathrm{ci}}$ are used to match the resonator into the readout line. In this configuration each detector within an array is probed by an individual readout tone at or near $f_i$. Figure 1(b) shows measured readout band transmission from one channel with $N$=132 detectors. The transmission minima represent the resonances. Figure 1(c) shows a simplified schematic of SAFE. During frame time $\tau_\mathrm{F}$ all $N$ detectors are addressed, each within time slot $\tau_\mathrm{S} = \tau_\mathrm{F}/N$. After passing through the detector array, the modulated RF signal is amplified and demodulated down to the baseband with a mixer using the unmodulated readout tone as the local oscillator.  The demodulation output contains the time-multiplexed detector signal, which is then analog-to-digital (A/D) converted at the sampling rate of $f_\mathrm{S} = 1/\tau_\mathrm{S}$.  In this approach, a benefit is that the high-speed digital electronics is limited to a controllable RF source with a sinusoidal output. The digital signal processing requirements are essentially reduced down to the data rate of $f_\mathrm{S}$ instead of the full readout band.

\subsection{Noise dynamics}

The risk in SAFE is a potential degradation of the signal-to-noise ratio (SNR) due to the limited integration time per detector which is a factor $1/N$ of the non-multiplexed continuous readout, or a fully parallel readout. For bolometers, however, the fundamental noise mechanism is the phonon noise \cite{mat1} which is characteristically band-limited white noise. Thus, if the frame rate $f_F$ is sufficiently faster than twice the thermal cutoff $f_c$ then, according to the Nyquist criterion, one obtains the full information on the baseband signal, and thus also avoids the noise penalty due to the aliasing effects. The low-pass thermal cutoff can be expressed as $f_{\mathrm{c}} = 1/(2\pi\tau_\mathrm{th})$. Here the thermal time constant is $\tau_\mathrm{th} = c/G$ with $c$ the heat capacity of the bolometer thermal volume, and $G$ the thermal conductivity from the volume to the thermal bath. In our practical devices $f_\mathrm{c}$ is in the range of tens of Hz up to somewhat over 100 Hz. In Appendix A1 we analyze the generalized case also accounting for the situation where the Nyquist criterion is not fully valid. Consequently, the single detector integration time (or $N$) has an impact on the SNR. We consider the case of an arbitrary frame rate. We also account for the contribution of a component of non-correlated wide-band white noise, which brings about the multiplexing penalty of $N^{1/2}$ as expressed in the root mean square (RMS) SNR.

In particular, the electronics noise needs to be accounted for in typical cases. To have a criterion for the electronics, we note that ideally the output from the detector, corresponding to the phonon noise, is expressed in voltage noise spectral density as $S_\mathrm{v,out} = (\Re\times\mathrm{NEP}_\mathrm{ph})^2$. Here $\Re$ is the voltage responsivity, and $\mathrm{NEP}_\mathrm{ph}$  is the phonon-noise limited noise equivalent power.  This is conveniently converted into equivalent noise temperature $T_\mathrm{e} = S_\mathrm{v,out}/(k_\mathrm{B} Z_0)$ with $k_\mathrm{B}$ the Boltzmann constant, and $Z_0$ the preamplifier input  impedance (typically $Z_0 = 50 \Omega$). This sets the criterion for the electronics noise temperature $T_\mathrm{N}<T_\mathrm{e}$, as referred to the first-stage preamplifier input. However, the electronics noise is typically uncorrelated in the time-scales of sampling. The corresponding noise elevation (see the last term of Eq. (A1) ) can be formally attributed  as an increase of the effective noise temperature of the preamplifier  into $NT_\mathrm{N}$. Thus, the criterion of the preamplifier noise in multiplexing becomes stricter, i.e., $T_\mathrm{N}<T_\mathrm{e}/N$. We note that in principle the responsivity in multiplexing can be made higher by increasing the excitation power.  This is possible, as the time averaged readout power is smaller in multiplexing as compared to the continuous readout of a single pixel. Thus, a higher excitation level can be used in the multiplexed readout increasing the responsivity, and consequently the noise temperature criterion does not necessarily strictly scale with $1/N$. 

 A further criterion for the readout scheme to function properly is that one needs to read out the detector signal within time slot $\tau_\mathrm{S}$. For KIBs this means recording the readout resonator state. Thus, the prerequisite is that the resonator is electrically sufficiently fast, i.e., the electrical time constant $\tau_\mathrm{el} = Q_\mathrm{t}/2\pi f_\mathrm{i}$ is shorter than $\tau_\mathrm{S}$. Here $Q_\mathrm{t}$ is the total quality factor of the KIB resonator including the intrinsic losses and those from the coupling to the readout line \cite{tim1}. We note that the criterion $\tau_\mathrm{S} > \tau_\mathrm{el}$ is equivalent to saying that the linewidth of the readout pulse  $\sim1/\tau_\mathrm{S}$ is narrow as compared to the linewidth of the resonance $ f_\mathrm{i}$. This also ensures that there will be no significant crosstalk effects between the detectors with adjacent resonant frequencies $f_\mathrm{i\pm 1}$ as the resonances are separated by several linewidths.

\section{Implementation} 
A block diagram of the electronics is shown in Fig. 2(a), along with the physical realization in Fig. 2(b). The system PC controls the microprocessor electronics, which is responsible of the real-time control of the readout sequence and A/D conversion. The microprocessor electronics controls the channel electronics block through serial peripheral interface (SPI) bus. In this implementation, we have one microprocessor controlling up to 13 readout channels. The channel electronics hosts the main functionalities of the readout sequence, i.e., RF excitation, RF preamplification, and demodulation. These are illustrated in Fig. 2(c). The RF excitation is performed by a direct digital synthesizer (DDS) chip (Analog Devices AD 9910) enabling the programming of the excitation sequences by defining the frequency for each time slot $\tau_\mathrm{S}$. The excitation signal first passes a digitally programmable attenuator, which defines its amplitude. After passing through the detector array the modulated RF signal is first amplified by a low-noise preamplifier. Thanks to the relatively high responsivity of the KIBs, the amplifier cascade can be entirely contained at room temperature, with the first-stage amplifier (Mini-Circuits PGA 103+) having a nominal noise temperature of 35 K. The modulated RF signal is then demodulated by an IQ mixer, generating the in-phase (I) and quadrature (Q) baseband components. The phase components contain the full information of the detector signal. However, for a given operating point, a single phase component with an optimal reference phase is sufficient for obtaining the detector signal. To halve the number of the required A/D channels, we have further implemented an analog summing circuit producing an output voltage proportional to the weighted sum of the voltages corresponding to the two phase components. As the relative weights are programmable, this process effectively corresponds to the post-selection of the phase reference. The output of the summing circuit contains the time-domain multiplexed (TDM) detector signals. To ensure the optimal integration of the signal with a negligible dead time, this signal is fed to a sample-and-hold (S/H) circuit \cite{hor1} integrating the signal across each time slot $\tau_\mathrm{S}$, and feeds the output to the A/D converter. The digitized TDM signal is then fed to the system PC responsible of demultiplexing and post-processing the data.
\begin{figure}
\begin{center}
\includegraphics[width =8cm]{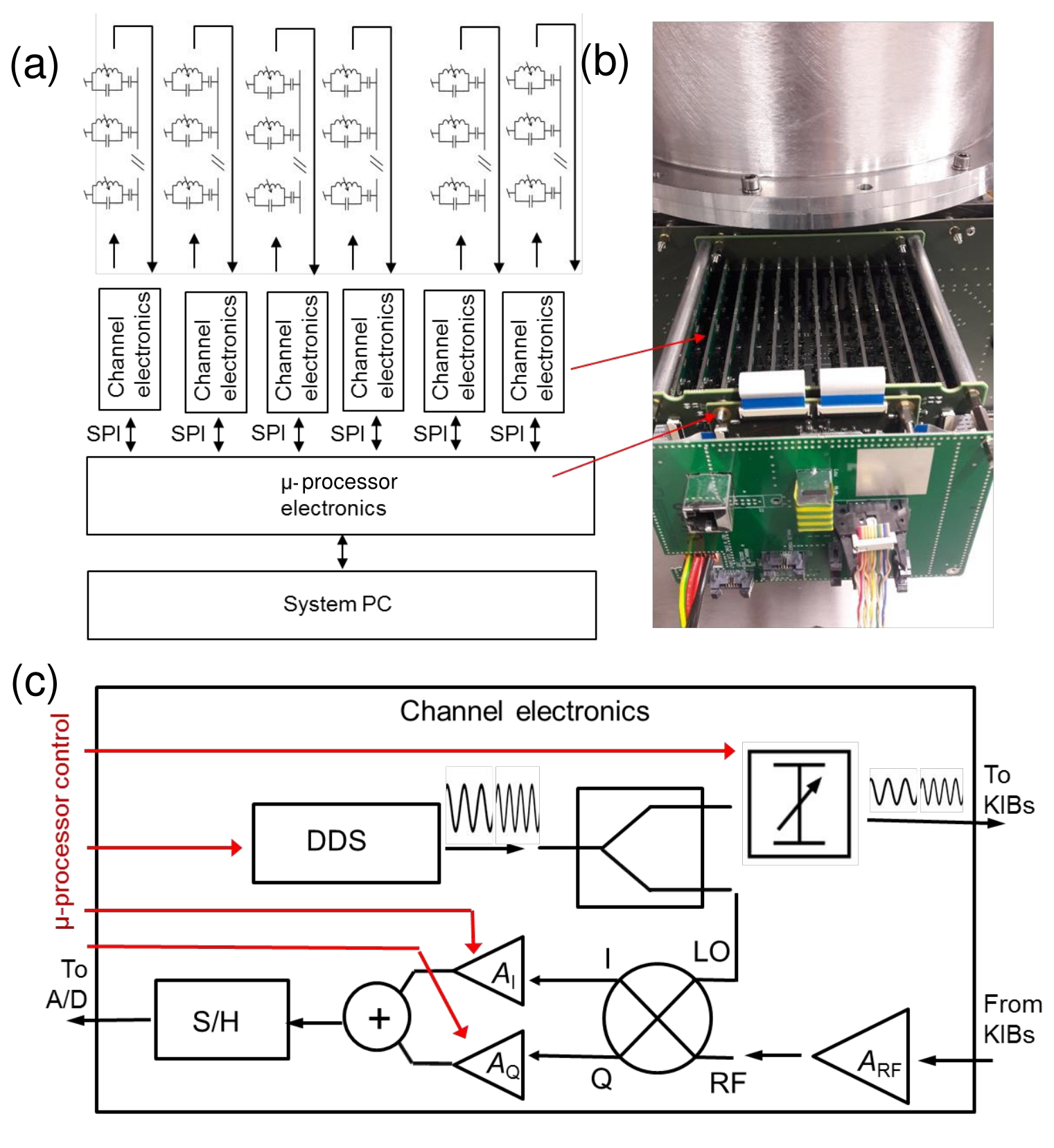}
\end{center}
\caption{(a) The electronics block diagram of the implementation of  SAFE and (b) a photograph of an electronics block installed onto ''LASTKID'' imaging system. (c) The signal flow in the channel electronics.}
\end{figure}

\section{Experiments}

\subsection{Imaging system}
We present the experiments demonstrating the functionality and properties of the multiplexing scheme with the SAFE electronics integrated into a KIB-based imaging system named "LASTKID". It is a fully staring imager with 8712 detector elements, and is currently under development. From the readout point of view, the LASTKID detector array is divided into 66 readout channels with $N=$132 detectors in each. We note that some previously reported experiments utilizing the KIB technology have been presented in the context of the ``CONSORTIS'' radiometer \cite{has1,gan1,dab1}. Here, we consider in particular the aspects of the readout with respect to the imager performance.

To overview the detector properties related to the readout, the phonon-limited noise equivalent power (NEP) is estimated as $\mathrm{NEP}_\mathrm{ph}=\sqrt{4 k_b T^2 G}$ from the thermal design ($G = 100$ nW/K, and operating temperature $T = 6.5$ K) as 15 fW/Hz$^{1/2}$. The responsivity depends on the inductance nonlinearity and the quality factor as $\Re = \sqrt{(\alpha/G)L^{-1}(dL/dT)Z_0 Q_i}$ in the limit of high intrinsic quality factor ($Q_i>Q_e$), where $Q_i$ and $Q_e$ are the intrinsic and coupling quality factors, respectively \cite{tim1}. We assumed here that we are observing the modulation of the quadrature phase component of the excitation signal which provides a monotonic response with the highest responsivity when the excitation tone frequency is chosen as $f_\mathrm{i}$. The constant $\alpha$ relates to the excitation power, and the stable operation requires $\alpha<0.8$. To be in the linear regime, somewhat lower value of $\alpha = 0.2$ is adopted for the estimates here.  From the data of the inset of Fig. 1b we extract the relevant measure of nonlinearity $L^{-1}(dL/dT)\approx$ 0.17 1/K. The coupling quality factor $Q_\mathrm{e}$ to the readout line is determined by the microwave design, i.e., by the selection of the coupling capacitors $C_\mathrm{ci}$ which are chosen to yield $Q_\mathrm{e}€$ between 80 and 180, somewhat varying along the readout line. The realized intrinsic quality factors $Q_i$ can be obtained by noting that the readout-band transmission at a resonant minimum is $S_{21}(f_\mathrm{i}) = Q/Q_\mathrm{i}$, where $Q=Q_\mathrm{e} Q_\mathrm{i}/(Q_\mathrm{e}+Q_\mathrm{i})$ is the total Q-value \cite{tim1}. From the data of Fig. 1(b) we can thus directly extract the quality factors in the range of 350 -- 2000. The reponsivities $\Re$ are thus estimated to be in the range of 0.8$\times 10^5$ V/W -- 1.8$\times 10^5$ V/W. The values of $\mathrm{NEP}$ and $\Re$ are in line with those obtained previously for similar detectors \cite{tim1,tim2}. Referring to the discussion in Section IIB, the readout electronics noise temperature criterion $T_N<(\Re\times\mathrm{NEP})^2/(k_\mathrm{B} Z_\mathrm{0} N)$ falls in the range of 16 K -- 80 K, i.e., in the order of the design value of the preamplifier $T_N$ = 35 K, but slightly below for some detectors. 

The electrical time constants $\tau_\mathrm{el}\approx Q_\mathrm{e}/(2\pi f_\mathrm{i})$ yield values below 1 $\mu$s for all the detectors. In the experiments the minimum slot time $\tau_\mathrm{S}$ is 60 $\mu$s, whence the criterion $\tau_\mathrm{e} <\tau_\mathrm{S}$ is well satisified.

We also briefly consider the linearity of the detectors. The resonant frequency shift of the $i$:th detector, due to the change of the detected signal $\Delta P$, is $\Delta f_\mathrm{i} = G^{-1}(df_i/dT)\Delta P$. With the relationship $f_\mathrm{i} = 1/(2\pi\sqrt{L(C_\mathrm{i}+C_\mathrm{ci}})$ this is readily written as $\Delta f_\mathrm{i}/f_\mathrm{i} = -(2G)^{-1}L^{-1}(\mathrm{d}L/\mathrm{d}T)\Delta P$. The dynamic range is approximately achieved by requiring that the relative frequency change $\Delta f_\mathrm{i}/f_\mathrm{i}$ is below the relative linewidth $1/Q$ of the resonance. The linear range is thus estimated as $\Delta P < 2GL(\mathrm{d}L/\mathrm{d}T)^{-1}/Q $. With the above numerical values this yields about $\Delta P \lesssim 7$ nW. The radiometric power variations from room temperature objects are expected to be well below 1 nW whence the linearity is considered sufficient.

\subsection{Readout tuning}

In a practical readout setting, a requirement is to find the correct RF excitation parameters, in particular the frequency and the phase reference, for each detector. The electronics output voltage $v_\mathrm{out}$ in a frequency sweep is proportional to the RF transmission in analogy to Fig. 1(b), albeit corresponding to just one phase component. To find the optimal frequency-dependent phase reference we first sweep the excitation frequency through the readout band. The transmission has a sinusoidal envelope due to the propagation delay of the excitation signal, which is depicted in Fig. 3(a). To fix the reference to a constant value at all frequencies, the phase reference post-selection circuit (see Fig. 2(c)) is programmed to compensate for the delay using the phase and frequency offset of the fitted envelope as the calibration. The excitation power level determining the responsivity was chosen in these measurements based on maximizing the excitation level while avoiding the emergence of the nonlinear effects due to the thermal feedback. This is detectable by the absence of the Duffing-like bending in the transmission characteristics \cite{tim1}. Consequently, the absolute value of the excitation was about 20 nW as referred to the input of the readout channel.  The transmission measurement is repeated with the corrected variable phase reference, and the result is plotted in Fig. 3(b). The responsivities of the detectors are proportional as $\Re \propto dv_\mathrm{out}/dL \propto f_\mathrm{i}\left(dv_\mathrm{out}/df\right)$. The corresponding responsivity spectrum, i.e., the frequency derivative of the data in Fig. 3(b) is plotted in Fig. 3(c), and the peak values are used as frequency calibration points $f_\mathrm{i}$. It can be noted that the responsivities of the individual detectors have some variation from detector to detector incident from the slight variation of the microwave matching conditions along the array. However, the responsivities as recorded here agree within a factor of $\sim 2$ which can be considered tolerable.
\begin{figure}[!t]
\begin{center}
\includegraphics[width = 12.5cm]{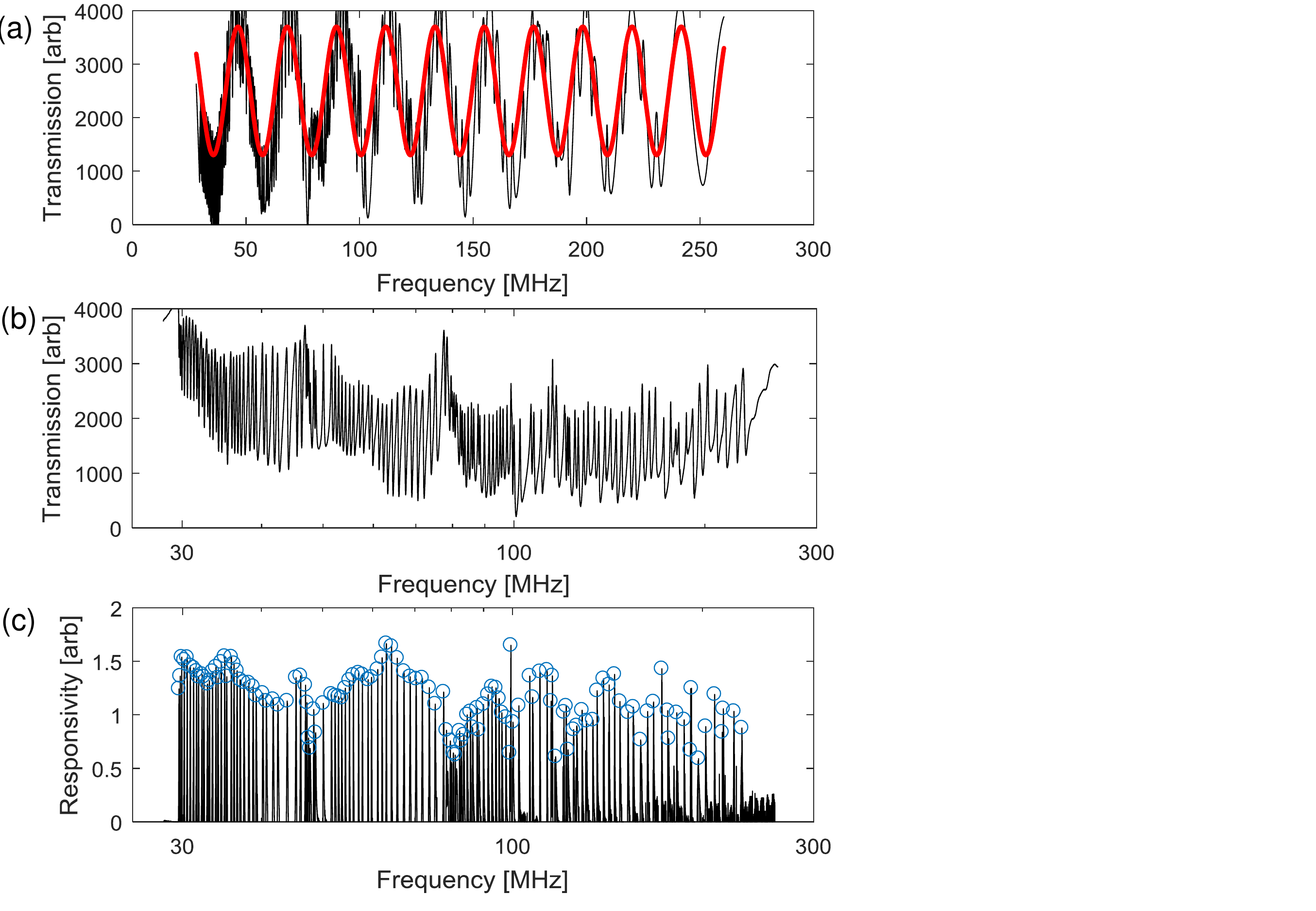}
\end{center}
\caption{(a) The transmission (proportional to detector output $v_\mathrm{out}$) of a KIB array as recorded by SAFE readout electronics with a constant phase reference.  The sinusoidal envelope corresponds to the propagation delay in the system. The thick red line is a fit to the envelope. (b) The transmission with the rotating phase reference as calibrated from the fit of (a). (c) The responsivities derived from the data of (b). The blue circles indicate the maxima corresponding to the optimal frequency operating points of the detectors.} 
\end{figure} 

\subsection{Signal-to-noise ratio in multiplexing}
To study the noise dynamics, we observe the effective SNR as the function of the multiplexing ratio $N$. For this end we utilize the radiometric calibration setup essentially similar to that previously used by us in an earlier prototype \cite{tim2}. The signal is stemming from an aqueous blackbody calibrator (ABC) \cite{die1} located at the optimal standoff distance of 2.5 m as defined by the quasioptics. The reference signal is that generated by applying an optical chopper in front of the ABC, with the chopping frequency of 42 Hz. The system-level SNR including the optics losses is quantified by the noise equivalent temperature difference (NETD), as SNR = $\Delta T/\mathrm{NETD}$, where $\Delta T$ is the temperature difference of the ABC and the background (room temperature). The initial calibration sweeping $\Delta T$ from room temperature up to about 40$^o$C in the absence of multiplexing yields NETD values between 37 mK/Hz$^{1/2}$ and 60 mK/Hz$^{1/2}$ for different pixels. In the case of multiplexing with $N$=114 detectors, and frame-rate $f_\mathrm{F}$=146 Hz, we obtain values between 110 mK/Hz$^{1/2}$ and 140 mK/Hz$^{1/2}$. This implies that there is some degradation of SNR due to the multiplexing, yet significantly below the $N^{1/2}$ worst-case limit. 

To study the effects to some more detail we take a closer look on the SNR data. It is readily observed that the signal level stays constant in multiplexing, as expected when the signal (chopper) frequency is below the Nyquist frequency of the sampling $f_\mathrm{F}/2$. Thus, it is sufficient to observe the behavior of the noise. We perform a set of measurements emulating the multiplexed noise dynamics, i.e., by limiting the signal integration time in post-processing to $\tau_\mathrm{F}/N$ in each frame of length $\tau_\mathrm{F}$. For verification, we test the equivalence of the emulation also with an actual multiplexing measurement. The data is presented in Fig. 4. We first note some nonidealities in the system: the detector thermal cutoff is about $f_\mathrm{c} = 120$ Hz as measured by a separate response measurement, whence the Nyquist criterion $f_\mathrm{F} > 2f_{c}$ is not exactly valid. This currently stems from certain electronics implementation issues preventing the increase of the frame rate. The theoretical data in Fig. 4 is plotted from Eq. (A1) by using the simulated correlated noise $S_\mathrm{BLW}(0,N)$ with $f_F/f_c \approx$1.2, and $\gamma$ as the fitting parameter. The extreme values of $\gamma$ are from 0.03 to 0.3. We note that we obtain the multiplexing penalties between 2.5 and 5.0 for $N=114$. For reference, the theoretical value with $\gamma$=0 is 1.6 with the given $f_F/f_c$. Yet, all measured noise values fall clearly below the multiplexing penalty $N^{1/2}$ as expected for the fully uncorrelated noise.
\begin{figure}[!b]
\begin{center}
\includegraphics[width = 9cm]{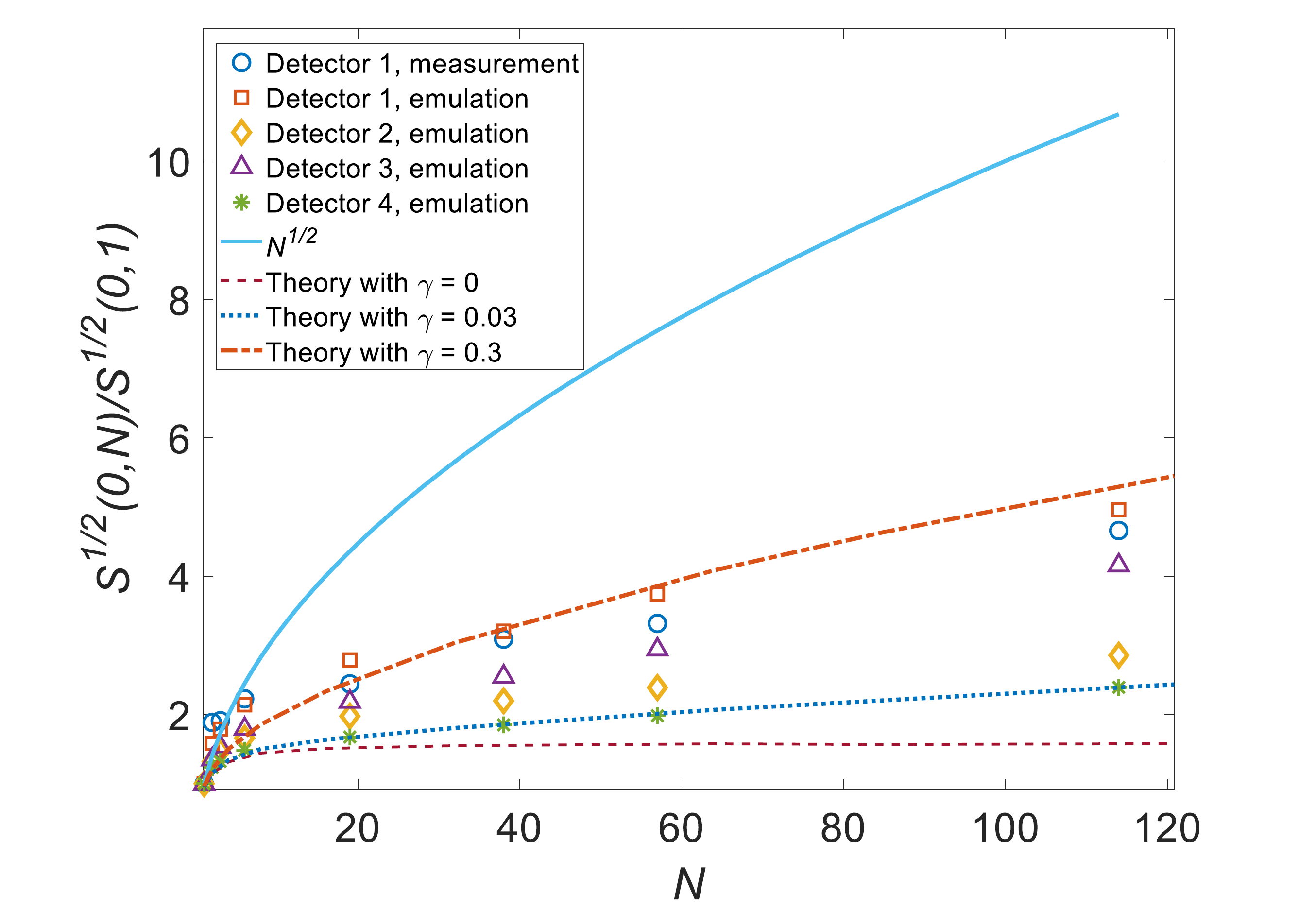}
\end{center}
\caption{The effective noise spectral density in the experiments either emulating the noise dynamics in multiplexing, or as obtained from an actual multiplexing experiment. The theoretical values are from Eq. (A1) with different $\gamma$. The worst-case noise penalty $N^{1/2}$ is shown as the solid  blue line.}
\end{figure} 

\subsection{Detection example}
Figure 5 illustrates a detection experiment using the SAFE readout. In Fig. 5(a) and (b) the focal plane with the detector area of 100 mm x 200 mm is shown. Currently, 2/3 of the detectors are mounted. Figure 5(c) shows a frame of data from a real-time data acquisition sequence obtained from a test scene where a person is concealing a plastic item with a lateral dimension of 100 mm and thickness of 5 mm in his pocket (Fig. 5d). The concealed object can be seen clearly as a notch at the field-of-view position $x_\mathrm{FOV}\approx$  0.4 m, providing contrast against the test person's skin (with the test person located at $x_\mathrm{FOV}$ between 0.2 m and 0.7 m), and thus demonstrating the capacity of the SAFE readout in the concealed object detection. The data was obtained along the line depicted in Fig. 5b, and is plotted as the effective radiometric temperature difference $\Delta T_\mathrm{rad}$. This was achieved by comparing the signal levels to those recorded with the ABC in the temperature range of 27--40 $^{o}$C. 
\begin{figure}[!t]
\begin{center}
\includegraphics[width = 9cm]{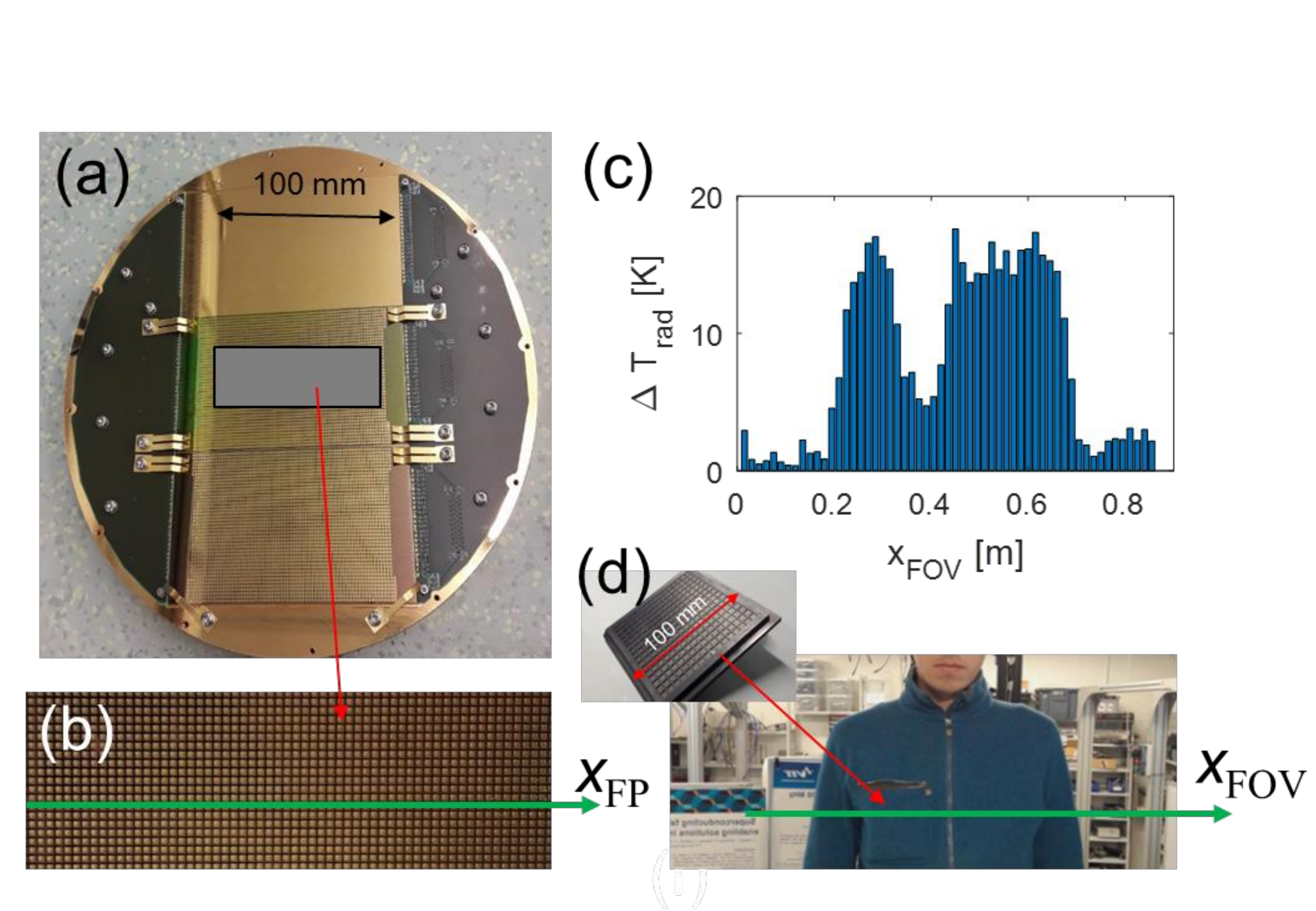}
\end{center}
\caption{(a)  The focal plane of the LASTKID system, and (b) a zoom-up of the detector matrix.  (c) A frame of data illustrating the detected radiometric power from a row of detectors presented with the post-integration time of 0.1 s. The data is calibrated into radiometric temperature difference $\Delta T_\mathrm{rad}$. (d) An optical photograph of the imaging situation, where a plastic object depicted in the inset is hidden in the pocket of the fleece jacket. The spatial coordinate system $x_\mathrm{FOV}$ corresponds to the focal plane coordinates $x_\mathrm{FP}$ referred to the detector matrix in (b) through the optics magnification.}
\end{figure}

\section{Discussion and conclusions}
In summary, we have presented a multiplexed electronics concept and an implementation designed  to read out kinetic inductance bolometer matrices. We indicated that the SAFE readout scheme presented here has beneficial features in comparison to other concepts relying on full FDM. In particular, the method allows a significant simplification as compared to multiplexers relying on high-speed digital electronics operating in the full readout frequency band. The constraints in SAFE are that, in order to avoid the multiplexing penalty, the detector responsivity and noise properties with respect to the sampling rate need to fulfill certain conditions which we reviewed in the context of KIBs. Beyond this, SAFE may be useful for the readout of other types of RF coupled sensors and detectors but the detector requirements need to be assessed for each case separately.

We discussed the criteria for avoiding the multiplexing penalty potentially degrading the SNR. This emphasizes the benefit of large detector arrays as an option to optomechanical scanning and a limited number of detectors: in the case of scanning the ``scanning penalty'' is always analogous to the worst-case multiplexing penalty, i.e. the RMS degradation of SNR is proportional to $N^{1/2}$. In this case $N$ represents the number of image pixels recorded with a single detector. We showed that this can be partially or completely avoided with large arrays and the SAFE multiplexing scheme. We also presented radiometric detection results from imaging systems indicating performance in practical real-time concealed object detection, and reported working progress towards a full-scale readout system for $\sim$ 9000 detectors. To date we have demonstrated the simultaneous readout of $>$ 500 detectors in radiometric detection.

\section*{Acknowledgement}
The authors would like to thank Markus Gr\"{o}nholm for the consultation about digital electronics and firmware, and Visa Vesterinen for useful discussions. The research received funding from European Space Agency (ESA) through contract GSTP 4000115091/15/NL/AF ``Large-area staring kinetic inductance focal plane array operating at elevated temperature'', and was also supported by European Union through FP7 project CONSORTIS, as well as  Academy of Finland through grants 305007, 314447, and 317844.

\appendix
\section{Multiplexing penalty for band-limited white noise}
To quantify the effect of the noise penalty including the nonidealities of the readout, we first numerically generate the band-limited white noise, and simulate the sampling process in our multiplexing scheme. The  noise is generated as a series of Gaussian distributed random numbers. This is subjected to the first-order numerical low-pass filter with cutoff $f_\mathrm{c}$ to impose the band limitation. The effect of multiplexing is simulated by dividing the time series in frames of length $\tau_\mathrm{F}$, and integrating the signal within each frame over a slot time $\tau_\mathrm{S} = \tau_\mathrm{F}/N$. The integrated values form a sequence with sampling rate $1/\tau_\mathrm{F}$, corresponding to the raw demultiplexed signal of a single detector. This is Fourier transformed, and the low-frequency limit $f \rightarrow 0$ of the RMS value of the power spectral density $S_\mathrm{BLW}(f,N)^{1/2}$ represents the effective noise in the multiplexing scheme with the multiplexing ratio of $N$. The result is plotted in Fig. 6 for different frame rates $f_\mathrm{F} = 1/\tau_\mathrm{F}$.
\begin{figure}
\begin{center}
\includegraphics[width = 8cm]{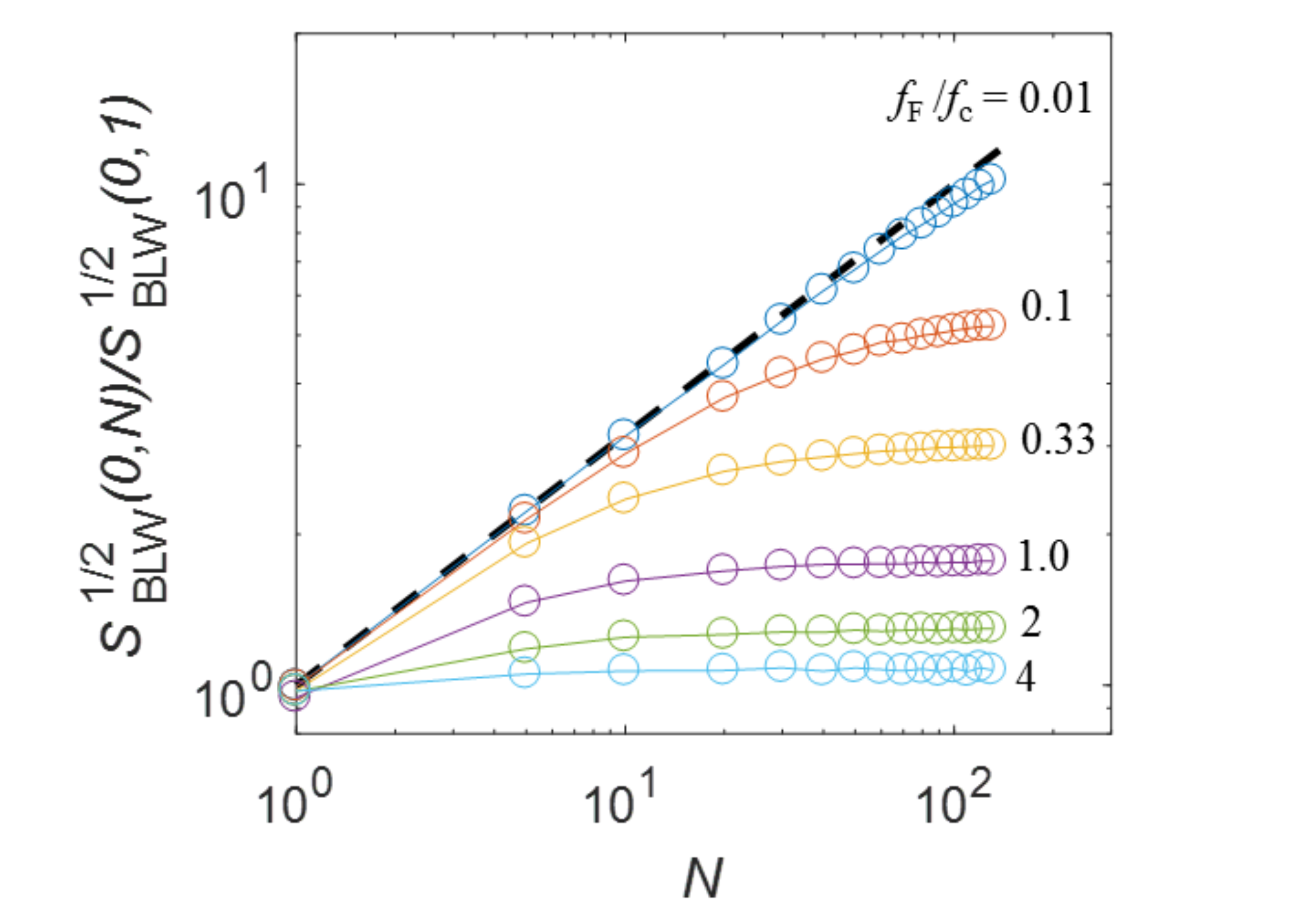}
\end{center}
\caption{Simulated low-frequency limit of effective root-mean-square noise power spectral density $S_\mathrm{BLW}(0,N)^{1/2} $ for $N$ detectors multiplexed with frame rate $f_\mathrm{F}$. It is assumed here that the dominant noise mechanism is band-limited white noise with high-frequency cutoff at $f_\mathrm{c}$. The results are normalized to the non-multiplexed case with $N=1$. The dashed line represents the worst-case scenario valid for non-correlated noise ($f_\mathrm{c} \rightarrow \infty$) proportional to $N^{1/2}$.}
\end{figure} 

To account for a component of uncorrelated noise (wide-band white noise), we express the total noise spectral density in the low-frequency limit as $S(0,N)=S_{\mathrm{BLW}}(0,N)+S_\mathrm{W}(0,N)$. Here, $S_\mathrm{BLW}(0,N)$ is as obtained from the simulations described above (Fig. 6), and $S_\mathrm{W}(0,N) = NS_\mathrm{W}(0,1)$ represents the uncorrelated noise suffering from the "full" multiplexing penalty. Expressing the non-multiplexed ratio of the noise components as $\gamma= S_\mathrm{W}(0,1)/S_\mathrm{BLW}(0,1)$ we can rewrite 
\begin{equation}
\frac{S(0,N)}{S(0,1)}=\frac{1}{\frac{S_\mathrm{BLW}(0,1)}{S_\mathrm{BLW}(0,N)}(1+\gamma)}+\frac{1}{\frac{1}{\gamma}+1}N.
\end{equation}


\begin{thebibliography}{33}

\bibitem{irw1} K. D. Irwin, and G. C. Hilton,  ``Transition edge sensors,'', in \textit{Cryogenic Particle Detection}, Springer, ed. Enss C, pp. 63-150 (2005). 

\bibitem{day1} P. L. Day, H. G. LeDuc, B. A. Mazin, A. Vayonakis, and J. Zmuidzinas,  Nature \textbf{425}, 817 (2003).

\bibitem{bas2} J. Baselmans, \textit{et al.}, J. Low Temp. Phys., \textbf{154}, 524 (2008).

\bibitem{hol1} W. S. Holland \textit{et al. }, Mon. Not. R. Astron. Soc, \textbf{430}, 2513 (2013).

\bibitem{bas1} J. J. A. Baselmans  et al., Astronomy" \& Astrophysics \textbf{601}, A89 (2017).

\bibitem{hei1} E. Heinz, \textit{et al.}, Opt. Eng., \textbf{50}, 113204 (2011).

\bibitem{row1} S. Rowe, \textit{et al.}, Rev. Sci. Instrum., \textbf{87}, 033105 (2016).

\bibitem{luu1} A. Luukanen, M. Gr\"onholm, M. M. Leivo, H. Toivanen, A. Rautiainen, J. Varis, Proc. SPIE \textbf{8362}, 836209 (2012).

\bibitem{luu2} A. Luukanen, T. Kiuru, M. M. Leivo, A. Rautiainen, and J. Varis, Proc. SPIE \textbf{8715}, 87150F (2013).

\bibitem{irw2} K. D. Irwin, Phys. C, \textbf{368}, 203 (2002).

\bibitem{hug1} S. McHugh, B. A. Mazin, B. Serfass, S. Meeker, K. O'Brian, R. Duan, R. Raffanti, D. Werthimer, Rev. Sci. Instrum. \textbf{83}, 044702 (2012).

\bibitem{tim1} A. V. Timofeev, V. Vesterinen, P. Helist\"o, L. Gr\"onberg, J. Hassel, J., and A. Luukanen, Supercond. Sci. Technol. \textbf{27}, 025002 (2014).

\bibitem{tim2} A. Timofeev \textit{et al}., IEEE Trans. THz Sci. and Technol., \textbf{7}, 218 (2017).

\bibitem{luo1} J. Luomahaara, V. Vesterinen, L. Gr\"onberg, and J. Hassel, Nature Commun., \textbf{5}, 4872 (2014).

\bibitem{mat1} J. C. Mather, Appl. Opt \textbf{21}, 1125 (1982).

\bibitem{hor1} P. Horowitz and W. Hill,  \textit{The art of electronics}, second. ed., Cambridge University Press (1989).

\bibitem{die1} C. Dietlein, Z. Popovic, and E. N. Grossman, Appl. Opt. \textbf{47}, 5604 (2008).

\bibitem{has1} J. Hassel et al., Proc. SPIE \textbf{10634}, 106340F (2018).

\bibitem{gan1} E. Gandini, A. Tamminen, A. Luukanen, and N. Llombart, IEEE Trans. on Antennas Propag., \textbf{66}, pp. 541 (2017).

\bibitem{dab1} Dabironezare S O, Hassel J, Gandini E, Gr\"{o}nberg L, Sipola H, Vesterinen V, and Llombart N, IEEE Trans. THz Sci. Technol. \textbf{8}, 746 (2018).

\end{thebibliography}
\end{document}